\documentstyle[12pt,aps,epsfig]{revtex}

\begin{document}

\baselineskip 24 pt

\title{Geometric Manipulation of Trapped Ions for Quantum Computation}
\author{L.-M.~Duan\thanks{
Email: Luming.Duan@uibk.ac.at}, J.I.~Cirac and P.~Zoller\\
Institute for Theoretical Physics, University of
Innsbruck, A-6020 Innsbruck, Austria \\
}
\maketitle

\begin{abstract}

We propose an experimentally feasible scheme to achieve quantum computation
based solely on geometric manipulations of a quantum system. The desired
geometric operations are obtained by driving the quantum system to undergo
appropriate adiabatic cyclic evolutions. Our implementation of the
all-geometric quantum computation is based on laser manipulation of a set of
trapped ions. An all-geometric approach, apart from its fundamental
interest, promises a possible way for robust quantum computation.
\end{abstract}

\baselineskip 24 pt

\bigskip

\bigskip

\newpage

The physical implementation of quantum computers requires a series of
accurately controllable quantum operations on a set of two-level systems
(qubits). These controllable quantum operations can be either of the
traditional dynamical origin ({\it 1}) or of a novel geometric origin ({\it
2, 3, 4, 5, 6, 7}). The all-geometric approach, proposed recently with the name
of holonomic quantum computation ({\it 4, 5, 6, 7}), achieves the whole set of
universal quantum gates solely based on the abelian and non-abelian
geometric operations (holonomies), without any contributions from dynamical
gates. The holonomies are acquired when a quantum system is driven to
undergo some appropriate cyclic evolutions by adiabatically changing the
controllable parameters in the governing Hamiltonian ({\it 8, 9, 10}).
The holonomies can be either simple abelian (commutable) phase factors (Berry
phases) or general non-abelian operations, depending on whether the
eigenspace of the governing Hamiltonian is nondegenerate or degenerate.
Besides its fundamental interest related to a general geometric global
structure, the holonomic quantum computation scheme has some built-in
fault-tolerant features ({\it 2, 7}), which might offer practical advantages,
such as being resilient to certain types of computational errors. Several
schemes have been proposed for the geometric realization of the particular
conditional phase shift gate with the use of the abelian Berry phase ({\it
2, 3}), and one of them has been experimentally demonstrated with the NMR
technique ({\it 2}). For a universal quantum computation, one still need to
combine this particular geometric gate together with some single-bit
dynamical gates ({\it 11}). Here, in contrast, we propose an
experimentally feasible scheme to achieve the universal quantum computation
all by the geometric means. This requires us to realize the non-abelian
holonomies as well as the abelian ones since the universal set of quantum
gates are necessarily non-commutable. Our scheme, which is based on laser
manipulation of a set of trapped ions, fulfils all the requirements of the
holonomic quantum computation and fits well the status of current technology.

For the holonomic quantum computation proposed recently ({\it 4, 5, 6, 7}), the
computational space $C$ is always an eigenspace (highly degenerate) of the
governing Hamiltonian, with a trivial eigenvalue $0$. Though the Hamiltonian
restricted to the computational space is completely trivial and there is no
dynamical evolution at all, the dependence of the Hamiltonian on some
controllable adiabatically changing parameters makes the space $C$ undergo a
highly nontrivial evolution due to the global geometric structure in the
parameter space. In fact, one requires any unitary operations in the space $%
C $ to be obtainable by these geometric evolutions in order to achieve
universal quantum computing. It is well known that some single-qubit
operations together with a non-trivial two-bit gate makes a universal set of
gate operations for quantum computing ({\it 11}). It is enough for us to
construct some looped paths in the parameter space to achieve the desired
geometric evolutions corresponding to these gate operations, and then a
composition of these parameter loops suffices to obtain an arbitrary unitary
evolution in the computational space $C$. In the following, we show how to
achieve all the desired geometric gate operations using a set of trapped
ions. The schemes for ion-trap quantum computers based on the conventionaldynamical
evolutions have been proposed ({\it 12, 13, 14, 15}), and some single-bit and
multi-bit gate operations have been demonstrated experimentally ({\it
16, 17, 18}). We use the same setup, but to achieve the holonomic
quantum computation. We also note that an idealized scheme ({\it 19}) was
proposed recently for holonomic quantum computation. Up to the best of our
knowledge, our proposal is the first realistic one which achieves all the
elements of holonomic quantum computation and is feasible with current
technology.

We choose the universal set of gate operations to be $U_{1}^{\left( j\right)
}=e^{i\phi _{1}\left| 1\right\rangle _{j}\left\langle 1\right| }$, $%
U_{2}^{\left( j\right) }=e^{i\phi _{2}\sigma _{j}^{y}}$, and $U_{3}^{\left(
jk\right) }=e^{i\phi _{3}\left| 11\right\rangle _{jk}\left\langle 11\right|
} $, where $\left| 0\right\rangle _{j}$ and $\left| 1\right\rangle _{j}$
constitute the computational basis for each qubit, $\sigma _{j}^{y}=i\left(
\left| 1\right\rangle _{j}\left\langle 0\right| -\left| 0\right\rangle
_{j}\left\langle 1\right| \right) $ is the Pauli operator of the $j$ qubit,
and $\phi _{1},\phi _{2},\phi _{3}$ are arbitrary phases. The universality
of this set of gates follows directly from the proof in ({\it 11}) and is
well known. First we show how to realize the single-bit gates $U_{1}^{\left(
j\right) }$ and $U_{2}^{\left( j\right) }$ geometrically. The system we have
in mind is a set of ions confined in a linear Paul trap ({\it 12, 17, 18}).
Each ion has three ground (or metastable) states $\left| 0\right\rangle
,\left| 1\right\rangle $ and $\left| a\right\rangle $, and one excited state
$\left| e\right\rangle $ (Fig.1). The state $\left| a\right\rangle $ is used
as an ancillary level for gate operations. The ground states could be
different hyperfine levels or in the same manifold but with different Zeeman
sublevels, and they are coupled to the excited state $\left| e\right\rangle $
separately by a resonant classical laser with a different polarization or
frequency (a possible separate addressing of the three levels is shown by
Fig. 1). The Hamiltonian for each ion with the laser on has the form
\begin{equation}
H_{j}=\hbar \left[ \left| e\right\rangle _{j}\left( \Omega _{0}\left\langle
0\right| +\Omega _{1}\left\langle 1\right| +\Omega _{a}\left\langle a\right|
\right) +\text{h.c.}\right]
\end{equation}
in the rotating frame, where $\Omega _{0},\Omega _{1},\Omega _{a}$ are Rabi
frequencies serving as the controlling parameters. Note that the Hamiltonian
in the rotating frame is independent of the laser frequencies since all the
lasers are resonant with the corresponding level transitions. The parameters
$\Omega _{0},\Omega _{1}$ should be set to zero initially so that the
computational space spanned by $\left| 0\right\rangle _{j}$ and $\left|
1\right\rangle _{j}$ is initially an eigenspace of the gate Hamiltonian with
a zero eigenvalue. Then the three Rabi frequencies make an adiabatic cyclic
evolution in the parameter space $M$ with the change rate significantly
smaller than the typical Rabi frequencies (the adiabatic condition), and the
adiabatic theorem assures that the computational space remains the
eigenspace of the gate Hamiltonian with the zero eigenvalue so there is no
dynamical phase contribution at all. However, we will explicitly show that
the topological holonomies accompanying the adiabatic evolutions suffice for
construction of the gates $U_{1}^{\left( j\right) }$ and $U_{2}^{\left(
j\right) }$, which in fact shows any single-bit\ operation is obtainable by
such holonomies.

To get the gate $U_{1}^{\left( j\right) }$, we set $\Omega _{0}=0$ so that
the state $\left| 0\right\rangle _{j}$ is decoupled, and choose $\Omega
_{1}=-\Omega \sin \frac{\theta }{2}e^{i\varphi }$, $\Omega _{a}=\Omega \cos
\frac{\theta }{2}$. The relative amplitude $\theta $ and phase $\varphi $ of
the Rabi frequencies $\Omega _{1}$ and $\Omega _{a}$ are the effective
control parameters, and the absolute magnitude $\Omega $ is irrelevant for
the gate control as long as it is large enough to satisfy the adiabatic
condition, which could be a good feature for real experiments. The dark
state (the eigenstate with the zero energy eigenvalue) of the gate
Hamiltonian has the form $\cos \frac{\theta }{2}\left| 1\right\rangle
_{j}+\sin \frac{\theta }{2}e^{i\varphi }\left| a\right\rangle _{j}$, where
the parameters $\theta ,\varphi $ make a cyclic evolution with the starting
and ending point to be $\theta =0$. Using the standard formula for the
geometric phase ({\it 8, 10}), we can show that this cyclic evolution achieves
the gate operation $U_{1}^{\left( j\right) }$ with the acquired Berry phase $%
\phi _{1}=\oint \sin \theta d\theta d\varphi $. This evolution has a
definite geometric interpretation: the acquired Berry phase is exactly the
enclosed solid angle $\oint d\Omega $ swept by the vector always pointing to
the $\left( \theta ,\varphi \right) $ direction. From this interpretation,
one immediately sees that the gate operation is only determined by the
global property, i.e., the swept solid angle, and does not depend on the
details of the evolution path in the parameter space. This is an advantage
of the holonomic quantum computation, which makes it robust against certain
types of errors. For instance, the local random errors along the evolution
path caused by some unwanted interaction would have very small influence on
the global property.

Now we show how to achieve the gate $U_{2}^{\left( j\right) }$
geometrically. For this purpose, we choose $\Omega _{0}=\Omega \sin \theta
\cos \varphi $, $\Omega _{1}=\Omega \sin \theta \sin \varphi $, and $\Omega
_{a}=\Omega \cos \theta $ in the Hamiltonian (1), with the parameters $%
\theta ,\varphi $ similarly undergoing an adiabatic cyclic evolution from $%
\theta =0$ to $\theta =0$.. The two degenerate dark states of this gate
Hamiltonian have the form $\left| D_{1}\right\rangle =\cos \theta \left(
\cos \varphi \left| 0\right\rangle _{j}+\sin \varphi \left| 1\right\rangle
_{j}\right) -\sin \theta \left| a\right\rangle _{j}$ and $\left|
D_{2}\right\rangle =\cos \varphi \left| 1\right\rangle _{j}-\sin \varphi
\left| 0\right\rangle _{j}$, from which we can show by using the formula for
holonomies ({\it 6, 9}) that the cyclic evolution of $\theta ,\varphi $
achieves the gate operation $U_{2}^{\left( j\right) }$ with the phase $\phi
_{2}=\oint d\Omega $, the swept solid angle by the vector $\left( \theta
,\varphi \right) $. The ability to obtain both of the non-commutable
geometric gates $U_{1}^{\left( j\right) }$ and $U_{2}^{\left( j\right) }$ in
fact shows that one constructs non-Abelian holonomies, since the composite
holonomies of the $U_{1}^{\left( j\right) }$ and $U_{2}^{\left( j\right) }$
and of the $U_{2}^{\left( j\right) }$ and $U_{1}^{\left( j\right) }$ are
different. Note that while the Abelian holonomies have been tested
experimentally by various means ({\it 10, 20}), the controllable
demonstration of the non-Abelian ones is believed to be more complicated
({\it 10, 19}). Here, in contrast, we introduce a simple way to test this
fundamental effect by manipulating a single ion with a laser. In fact, for
the demonstration of the non-Abelian holonomies we do not need to exploit
any interaction between the ions, so one can also use a sample of free
particles instead of a single ion for a simple test. For instance, one can
experimentally verify this by laser manipulation of a cloud of atoms in a
magnetic-optical trap, which is readily available in many laboratories.

A combination of the gates $U_{1}^{\left( j\right) }$ and $U_{2}^{\left(
j\right) }$ permits to implement any single-bit operation, which together
with the nontrivial two-bit gate $U_{3}^{\left( jk\right) }$ between the
qubits $j,k$, are enough for universal quantum computation. To construct the
gate $U_{3}^{\left( jk\right) }$ using geometric means, we need exploit the
Coulomb interactions between the ions. For this purpose, we provide a scheme
based on the recent dynamical proposal ({\it 14}) which uses two-color laser
manipulation. The transition $\left| 1\right\rangle \rightarrow \left|
e\right\rangle $ for the $j,k$ ions is driven by a red and a blue detuned
laser, respectively with detunings $-\left( \nu +\delta \right) $ and $\nu
+\delta $ (Fig. 2), where $\nu $ is the phonon frequency of one oscillation
mode (normally the center of mass mode) and $\delta $ is an additional
detuning. Similarly, the transition $\left| a\right\rangle \rightarrow
\left| e\right\rangle $ is also driven by a red and a blue detuned laser,
but with the additional detuning $\delta ^{\prime }\neq \delta $ to avoid
the direct Raman transition. For simplicity here we choose $\delta ^{\prime
}=-\delta $ as in Fig. 2. Under the condition of strong confinement $\eta
^{2}\ll 1$ (the Lamb-Dicke criterion), where $\eta $ is defined by the ratio
of the ion oscillation amplitude to the manipulation optical wave length,
the Hamiltonian describing the interaction has the form
\begin{equation}
H_{jk}=\frac{\eta ^{2}}{\delta }\left[ -\left| \Omega _{1}\right| ^{2}\sigma
_{j1}^{\varphi _{1}}\sigma _{k1}^{\varphi _{1}}+\left| \Omega _{a}\right|
^{2}\sigma _{ja}^{\varphi _{a}}\sigma _{ka}^{\varphi _{a}}\right] ,
\end{equation}
where $\sigma _{j\mu }^{\varphi _{\mu }}\equiv e^{i\varphi _{\mu }}\left|
e\right\rangle _{j}\left\langle \mu \right| +$h.c. $\left( \mu =1,a\right) $
and $\Omega _{1},\Omega _{a}$ are the corresponding Rabi frequencies
respectively with the phases $\varphi _{1},\varphi _{a}$. In writing the
Hamiltonian (2) we have neglected some trivial light shift terms which can
be easily compensated, for instance, by another laser. To get a geometric
operation, we choose the relative intensity $\left| \Omega _{1}\right|
^{2}/\left| \Omega _{a}\right| ^{2}=\tan \left( \theta /2\right) $ and phase
$\varphi _{1}-\varphi _{a}=\varphi /2$, with the control parameters $\theta
,\varphi $ undergoing a cyclic adiabatic evolution from $\theta =0$. During
the evolution, the computational bases $\left| 00\right\rangle _{jk},\left|
01\right\rangle _{jk}$ and $\left| 10\right\rangle _{jk}$ are decoupled from
the Hamiltonian (2), while the $\left| 11\right\rangle _{jk}$ component
adiabatically follows as $\cos \frac{\theta }{2}\left| 11\right\rangle
_{jk}+\sin \frac{\theta }{2}e^{i\varphi }\left| aa\right\rangle _{jk}$,
which acquires a Berry phase after the whole loop. So we get the conditional
phase shift gate $U_{3}^{\left( jk\right) }$ with the purely geometric phase
$\phi _{3}=\oint d\Omega $, the swept solid angle by the vector $\left(
\theta ,\varphi \right) $. Note that this geometric two-bit gate has shared
the advantages of the recently proposed and demonstrated dynamical scheme
({\it 14, 18}) in the sense that: first, the ion motional modes need not be
cooled to their ground states as long as the Lamb-Dicke criterion is
satisfied; and second, we do not need separate addressing of the ions during
the two-bit gate operation.

For experimental demonstration of the above universal set of geometric
gates, we need to consider several kinds of decoherence which impose
concrete conditions on the relevant parameters. Firstly, one should fulfil
the adiabatic condition. This means the gate operation time should be larger
than the inverse of the energy gap between the dark states and the bright
and excited states. The energy gap is given by $\Delta _{1}=\left| \Omega
\right| $ for the single-bit gates and by $\Delta _{2}=\eta ^{2}\left|
\Omega \right| ^{2}/\delta $ for the two-bit gate. So we require the
single-bit and the two-bit gate operation times $t_{i}^{g}$ $\left(
i=1,2\right) $ are reasonably long so that the leakage error to the bright
and the excited states, which scales as $1/\left( \Delta
_{i}t_{i}^{g}\right) ^{2}$, is small. Secondly,\ we need to avoid
spontaneous emission (with a rate $\gamma _{s}$) of the excited state $%
\left| e\right\rangle $. Due to the adiabatic condition the excited state is
only weakly populated even though we use resonant laser coupling, and the
effective spontaneous emission rate is reduced by the leakage probability $%
1/\left( \Delta _{i}t_{i}^{g}\right) ^{2}$. As a result we only require $%
\gamma _{s}/\left( \Delta _{i}^{2}t_{i}^{g}\right) \ll 1$ for the
spontaneous emission to be negligible during the gate operation. Finally,
the influence of the heating of the ion motion should be small. We assume
all the manipulation lasers are copropagating so that the heating caused by
the two-photon recoils is negligible. The carrier phononic states are only
virtually excited during the two-bit gate, so the influence of the heating
rate $\gamma _{h}$ is reduced by the phonon population probability $\eta
^{2}\left| \Omega \right| ^{2}/\delta ^{2}$. The effective heating rate
should be much smaller than the gate speed, which requires $\delta \gg $ $%
\gamma _{h} \ll \delta$. Note that all the conditions discussed above in the geometric
gates are exactly parallel to those in the dynamical schemes using the off
resonant Raman transitions. The reason for this is that in both cases the
excited state is only weakly populated and the population probability obeys
the same scaling law, though the physical mechanism for the weak population
is quite different. So, compared with the dynamical schemes, our
requirements are not more stringent. However, the geometric proposal
introduced in this paper,\ on the one hand, will permit us to investigate
experimentally the fundamental Abelian and non-Abelian holonomies ({\it 10})
, and on the other hand, may open new possibilities for robust quantum
computation ({\it 21, 22}).

Note added: After submission of this work, we became aware that a
non-abelian holonomy has been designed for atoms by Unanyan,
Shore, and Bergmann [Phys. Rev. A 59, 2910 (1999)].

{\bf Caption of Fig. 1:}

Level structure and laser configuration for single-bit operations.
A possible choice for the three ground or metastable states is that $\left|
1\right\rangle $ and $\left| a\right\rangle $ are two degenerate Zeeman
sublevels which are addressed by lasers with different polarizations, and $%
\left| 0\right\rangle $ is the ground state (another hyperfine level) with slightly different
energy so that it can be addressed by a laser with a different frequency.

{\bf Caption of Fig. 2:}

Laser configuration for the two-bit operation. The same
configuration for both ions.

\newpage
\begin{figure}[tbp]
\epsfig{file=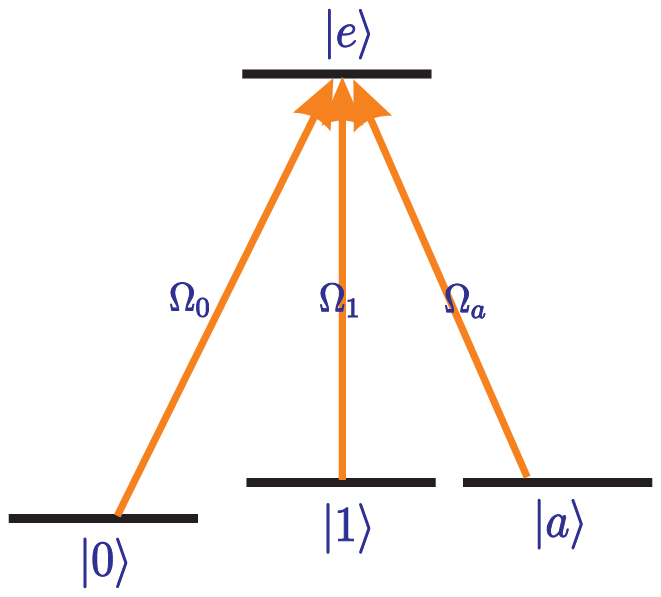,width=8cm} \caption{}
\end{figure}
\begin{figure}[tbp]
\epsfig{file=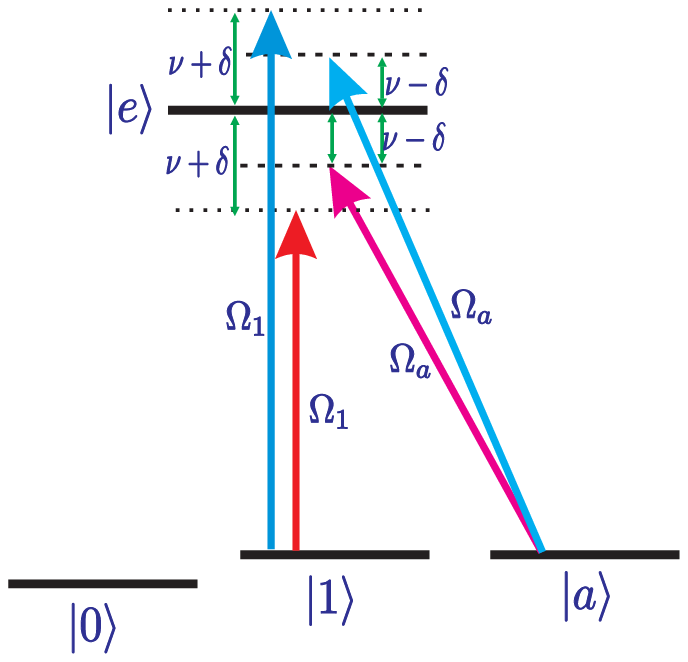,width=8cm} \caption{}
\end{figure}

\end{document}